\documentclass[journal]{IEEEtran}
\usepackage{amsmath,amsfonts}
\usepackage{algorithmic}
\usepackage{algorithm}
\usepackage{array}
\usepackage[caption=false,font=normalsize,labelfont=sf,textfont=sf]{subfig}
\usepackage{textcomp}
\usepackage{stfloats}
\usepackage{url}
\usepackage{verbatim}
\usepackage{graphicx}
\usepackage{cite}
\usepackage{multirow}
\usepackage{color}
\usepackage{makecell}
\begin{document}

\title{3DAttGAN: A 3D Attention-based Generative Adversarial Network for Joint Space-Time Video Super-Resolution}

\author{Congrui Fu, Hui Yuan, Liquan Shen, Raouf Hamzaoui, and Hao Zhang
\thanks{Corresponding author: Hui Yuan (e-mail: huiyuan@sdu.edu.cn).}
}



\maketitle

\begin{abstract}
\textcolor{black}{
In many applications, including surveillance, entertainment, and restoration, there is a need to increase both the spatial resolution and the frame rate of a video sequence. The aim is to improve visual quality, refine details, and create a more realistic viewing experience. Existing space-time video super-resolution methods do not effectively use spatio-temporal information.
To address this limitation, we propose a generative adversarial network for joint space-time video super-resolution.
The generative network consists of three operations: shallow feature extraction, deep feature extraction, and reconstruction. 
It uses three-dimensional (3D) convolutions to process temporal and spatial information simultaneously and includes a novel 3D attention mechanism to extract the most important channel and spatial information.    
The discriminative network uses a two-branch structure to handle details and motion information, making the generated results more accurate. Experimental results on the Vid4, Vimeo-90K, and REDS datasets demonstrate the effectiveness of the proposed method.
The source code is publicly available at https://github.com/FCongRui/3DAttGan.git.
}
\end{abstract}

\begin{IEEEkeywords}
video space-time super-resolution, generative adversarial network, 3D attention mechanism, quality enhancement.
\end{IEEEkeywords}

\section{Introduction}

Video super resolution (VSR) aims at estimating high-resolution (HR) frames from low-resolution (LR) ones. VSR has many applications, including face recognition~\cite{2}, small object recognition~\cite{1}, and 4K ultra-high-definition (UHD) video~\cite{3}.
\textcolor{black}{
Usually, VSR refers to spatial super-resolution (SSR).
The aim is to generate an HR video sequence 
from a corresponding LR video sequence. 
To achieve SSR, two main types of methods are used: model-based methods and learning-based methods. 
In model-based SSR methods, e.g.,~\cite{4,5,6,7}, the LR frames are modeled as blurred and downsampled versions of the HR frames.}
Using this model, it is possible to estimate the HR frames with an inverse calculation.
However, as the SSR problem is ill posed, regularization techniques are necessary to effectively reconstruct the HR frames.
To incorporate specific image features into the HR estimate, prior information is often used. 
For instance, in the Bayesian framework, the SSR problem can be statistically modeled and regularized with the integration of smoothness and sparsity priors.
In contrast to model-based schemes, learning-based schemes such as deep neural networks (DNNs) do not derive an analytical SSR model. Instead, they leverage large training datasets containing both HR and LR videos to learn how to solve the VSR problem.

\begin{figure*}[htbp]
	\centering
	\includegraphics [width=7in]{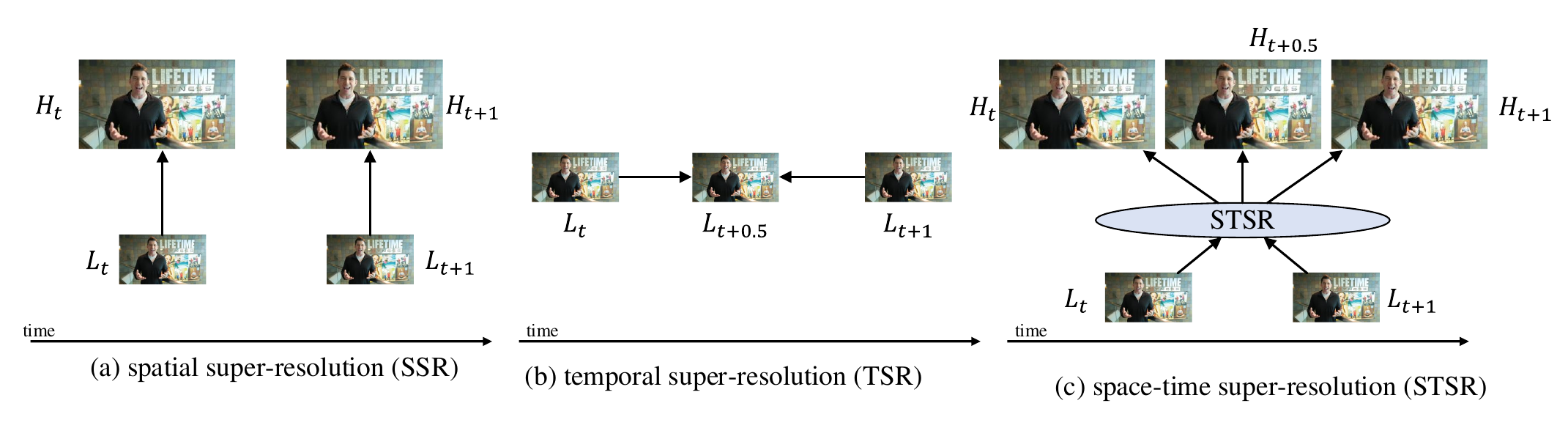}
	\caption{Illustrations of three kinds for VSR: (a) SSR, (b) TSR, and (c) STSR. $ H $  denotes a high spatial resolution frame and $ L $ denotes a low spatial resolution frame. The subscripts $ t $, $ t+0.5 $, and $ t+1 $ denote time indices. }
	\label{VSR}
\end{figure*}

Besides SSR, there are two other kinds of VSR: temporal super-resolution (TSR) and space-time super-resolution (STSR). In TSR, a video frame is interpolated between existing video frames (Fig.~\ref{VSR}(b)). On the other hand, in STSR, a video with high space-time resolution is generated from a given video with low space-time resolution (Fig.~\ref{VSR}(c)). In this paper, we focus on STSR. 

Existing STSR methods usually achieve spatial and temporal super-resolution independently by applying TSR and SSR alternately. This approach is not only time consuming but also inefficient as it is difficult to take advantage of the high temporal resolution if SSR is applied first and to take advantage of the high spatial resolution if TSR is applied first. A higher spatial representation can help improve motion estimation accuracy, which is important for TSR, while a higher temporal resolution can enhance the accuracy of SSR by capturing finer details in successive frames due to their similarity.

\textcolor{black}{
To harness the potential of the spatial and temporal features for VSR, we propose a generative adversarial neural network (GAN) with a 3D attention mechanism that simultaneously generates high-spatial-resolution and high-frame-rate video frames from low-spatial-resolution and low-frame-rate input video frames.
}

We use 3D convolution operations to process temporal and spatial information simultaneously and extend the standard 2D attention mechanism to 3D to make it suitable for STSR. Our generative network consists of shallow feature extraction, deep feature extraction, and reconstruction. Multiple residual attention blocks with a 3D attention mechanism are superimposed to enhance the accuracy of spatial and temporal feature extraction. The discriminative network is used to distinguish between the outputs of STSR and real HR videos. It includes two branches; one evaluates details, and the other assesses temporal information using motion consistency analysis. 
The contributions of the paper can be summarized as follows. 

\textcolor{black}{
\begin{itemize}     
	\item[$\bullet$] We propose a novel generative adversarial network to simultaneously enhance the spatial and temporal resolution of input videos. 
	Our network has less complexity than state-of-the-art STSR methods and outperforms them in areas of rich texture and high motion. 
	\item[$\bullet$] We propose a 3D attention mechanism that uses 3D convolutions to extract important temporal and spatial information simultaneously. Current STSR attention mechanisms are designed for 2D convolutional networks and cannot be directly applied to 3D convolutions due to inconsistent dimensions. 
	\item[$\bullet$] We design two discriminator strategies that significantly enhance the performance of the proposed generator. One strategy focuses on the spatial detail characteristics of video frames, while the other emphasizes the temporal motion information. This is the first time that a two-branch discriminator is proposed for video STSR. 	
\end{itemize}
}

The remainder of this paper is organized as follows. Section II gives an overview of learning-based VSR. Section III provides a detailed description of the proposed network. Experimental results, including an ablation study, are presented in Section IV. Finally, Section V provides a summary of the paper.

\section{Related work}

\subsection{Video spatial super-resolution}

Video SSR improves the quality of low-resolution videos by upscaling them to higher resolutions.
The methods based on deep learning use external references to construct models while simultaneously approximating complex nonlinear functions.
SRCNN~\cite{12}, one of the pioneering convolutional neural networks (CNN), is a simple three-layer end-to-end model.
Video SSR methods take advantage of the temporal correlations by merging spatial information over adjacent frames and estimate the inter-frame motion to obtain references.
Unlike single image super-resolution, VSR can exploit motion information between consecutive frames to recover lost high-frequency details in the current frame.

Tian \textit{et al.}~\cite{13} proposed a temporal deformable alignment network (TDAN) that can align the reference image and supporting frames without requiring optical flow estimation.
Wang \textit{et al.}~\cite{14} proposed a video super-resolution network, EDVR, that uses enhanced deformable convolutions. EDVR comprises a pyramid, cascading, and deformable (PCD) alignment module to effectively address the motions, as well as a temporal and spatial attention (TSA) fusion module that emphasizes crucial features for subsequent restoration. 
Chan \textit{et al.}~\cite{15} proposed a concise pipeline for VSR called BasicVSR, which leverages bidirectional propagation to enhance information gathering, and uses a method based on optical flow to estimate the correspondence between adjacent frames for accurate feature alignment.
Building on SRGAN~\cite{22}, which uses a modified version of the VGG~\cite{23} network as the discriminator, ESRGAN~\cite{24} incorporates a residual-in-residual dense block and uses a relativistic average GAN (RaGAN)~\cite{25} to evaluate the authenticity of images.
VSRResNet~\cite{VSRResNet} relies on a discriminator architecture during GAN training and uses two regularizers: a feature-space loss and a pixel-space loss. 
For video super-resolution, temporally coherent GAN (TecoGAN)~\cite{27} was the first method to introduce a spatiotemporal discriminator that takes into account temporal and spatial coherence. 
TecoGAN is based on frame recurrent video super resolution (FRVSR)~\cite{26}, which uses the HR estimate from the preceding frame for super-resolving the next frame.
\textcolor{black}{
Wen \textit{et al.}~\cite{s4} proposed an end-to-end deep convolutional network for video super-resolution that dynamically generates spatially adaptive filters to improve temporal alignment.
Yi \textit{et al.}~\cite{s5} introduced a progressive fusion network for video super-resolution that effectively fuses temporal information, incorporates multi-scale structures, hybrid convolutions, and non-local operations, improving both performance and complexity.
Li \textit{et al.}~\cite{s1} presented a novel approach for efficient video resolution upscaling by leveraging spatial-temporal information to divide videos into chunks, reducing model size and storage requirements while achieving real-time video super-resolution with high quality.
Qiu \textit{et al.}~\cite{s2} introduced Frequency-Transformer for Video Super-Resolution (FTVSR++). The method operates in a combined space-time-frequency domain, distinguishes real visual texture from artifacts, and incorporates dual frequency attention.
Meng \textit{et al.}~\cite{s3} introduced an approach to enhance video super-resolution by leveraging similar patches across distant frames through long-term cross-scale aggregation. The method is adaptable as post-processing for any super-resolution technique.
}

\subsection{Video temporal super-resolution}

Video TSR mitigates motion artifacts and reduces judder by interpolating frames between existing ones, proving valuable for slow-motion effects creation and enhancing overall video playback quality.
Video TSR is similar to video frame interpolation (VFI). The objective is to predict one or multiple frames between two input frames. While traditional methods typically use dense optical flow interpolation, recent methods often use DNNs for middle frame prediction.

Liu \textit{et al.}.~\cite{16} proposed a deep voxel flow (DVF) network, which combines the advantages of classic flow-based methods with those of learning-based methods. 
Bao \textit{et al.}~\cite{17} introduced a network that uses an adaptive warping layer to process video data through a combination of optical flow-based warping and learned interpolation kernel-based sampling of input frames.
Xue \textit{et al.}~\cite{34} presented ToFlow, a tailored flow representation for video processing tasks, achieved through a neural network incorporating motion estimation and video processing components. 
Bao \textit{et al.}~\cite{18} introduced a VFI approach that generates more realistic intermediate flows by prioritizing closer objects over distant ones using a depth-aware flow projection layer.
%
The cycle consistency network proposed by Liu \textit{et al.}~\cite{19} measures the quality of synthesized frames by evaluating their ability to reconstruct the input frames accurately, thus increasing their reliability.
Park \textit{et al.}~\cite{20} proposed a VFI method that identifies and exploits the position and strength of complicated motion to generate intermediate frames.
Lee \textit{et al.}~\cite{21} proposed a method called ``adaptive collaboration of flows (AdaCoF)'' that can generate a target pixel by referring to a variable number of pixels from any position.
\textcolor{black}{
Kong \textit{et al.}~\cite{t1} introduced a Progressive Motion Context Refine Network for efficient frame interpolation. The network predicts motion fields and image context jointly, simplifying the task by reusing existing textures from adjacent input frames. It achieves favorable results with a reduced model size and running time.
Liu \textit{et al.}~\cite{t2} proposed a novel trajectory-aware transformer for video frame interpolation. The method formulates the warped features with inconsistent motions as query tokens and uses relevant regions in a motion trajectory from two original consecutive frames as keys and values.
Park \textit{et al.}~\cite{t3} introduced BiFormer, a 4K video frame interpolator, which uses a bilateral transformer to predict global motion fields, refines them using blockwise bilateral cost volumes, and synthesizes intermediate frames.
Plack \textit{et al.}~\cite{t4} presented a novel transformer-based video frame interpolation network that estimates both the interpolated frame and its expected error. The method improves visual quality, provides error maps for identifying problematic frames, and supports partial rendering passes to enhance frame quality while reducing processing time.
Zhu \textit{et al.}~\cite{t5} presented MFNet, a frame interpolation network that focuses on motion regions and uses methods such as adaptive motion region separation, fine-grained approximation of intermediate streams, and lightweight bi-directional optical stream fusion.
}

\subsection{Video space-time super-resolution}  

\textcolor{black}{ 
Video STSR enhances both spatial and temporal resolutions.
Video STSR is a very challenging inverse problem as LR frames lack detailed texture and motion information.  
The common approach to STSR for videos is to alternate between SSR and TSR. 
However, this approach leads to excessive complexity. 
A better approach is to generate videos that have both HR and high-frame-rate simultaneously.
However, existing methods do not fully exploit the temporal and spatial characteristics across video frames. 
}

Shechtman \textit{et al.}~\cite{53} proposed a method that creates an STSR video of a scene by combining information from a number of LR videos of the same scene. The method is based on modeling each LR video as a blurred version of the HR video in space and time. 
%
Mudenagudi \textit{et al.}~\cite{54} formulated video STSR as a reconstruction problem in a Markov random field-maximum a posteriori framework and used graph-cut optimization to solve the problem. 
%
Takeda \textit{et al.}~\cite{57} exploited MASK regression, a method that uses local spatial orientations and motion vectors to construct adaptive filters at every position of interest.
Shahar \textit{et al.}~\cite{58} combined information from multiple space-time patches to super resolve input videos. 
Haris \textit{et al.}~\cite{STSR} developed a CNN that super-resolves video jointly in space and time, incorporating direct lateral connections between multiple resolutions to present multi-scale features during training. 
Kang \textit{et al.}~\cite{b1} proposed a weighting scheme for efficient video processing, where all input frames are fused without the need for explicit motion compensation.
Dutta \textit{et al.}~\cite{b2} used quadratic modeling for LR interpolation and reused flowmaps and blending masks for both LR and HR synthesis.
Xiang \textit{et al.}~\cite{b3} proposed a deformable feature interpolation network to capture local temporal characteristics when interpolating LR frame features. They also introduced a deformable ConvLSTM to align and gather time information, leading to improved utilization of global temporal contexts.
Xu \textit{et al.}~\cite{b5} introduced a method that uses a temporal modulation block (TMB) to control feature interpolation and a locally-temporal feature comparison (LFC) module to capture time-based information in video processing.
\textcolor{black}{
Zhang \textit{et al.}~\cite{st1} proposed cross-frame transformers instead of traditional convolutions divided the input feature sequence into query, key, and value matrices,along with a multi-level residual reconstruction module. This approach allows for the use of the maximum similarity and similarity coefficient matrices obtained though the cross-frame transformer.
}


\section{Proposed method}

Given a low resolution video $\textit{\textbf{V}}^{in}= \left\{\textbf{x}_{1}^{low }, \textbf{x}_{2}^ {low },  \ldots, \textbf{x}_{n}^{low }\right\}$ with $ n $ frames, the STSR task is to generate the corresponding HR video frames $\textit{\textbf{V}}^{out}= \left\{\textbf{y}_{1}^{high}, \textbf{y}_{2}^{high },  \ldots, \textbf{y}_{2 n-1}^{high }\right\}$ with $ 2n-1 $ frames, where the superscripts ``\textit{low}'' and ``\textit{high}'' denote spatial resolutions.

\begin{figure*}[htbp]
	\centering
	\includegraphics [width=5.5in]{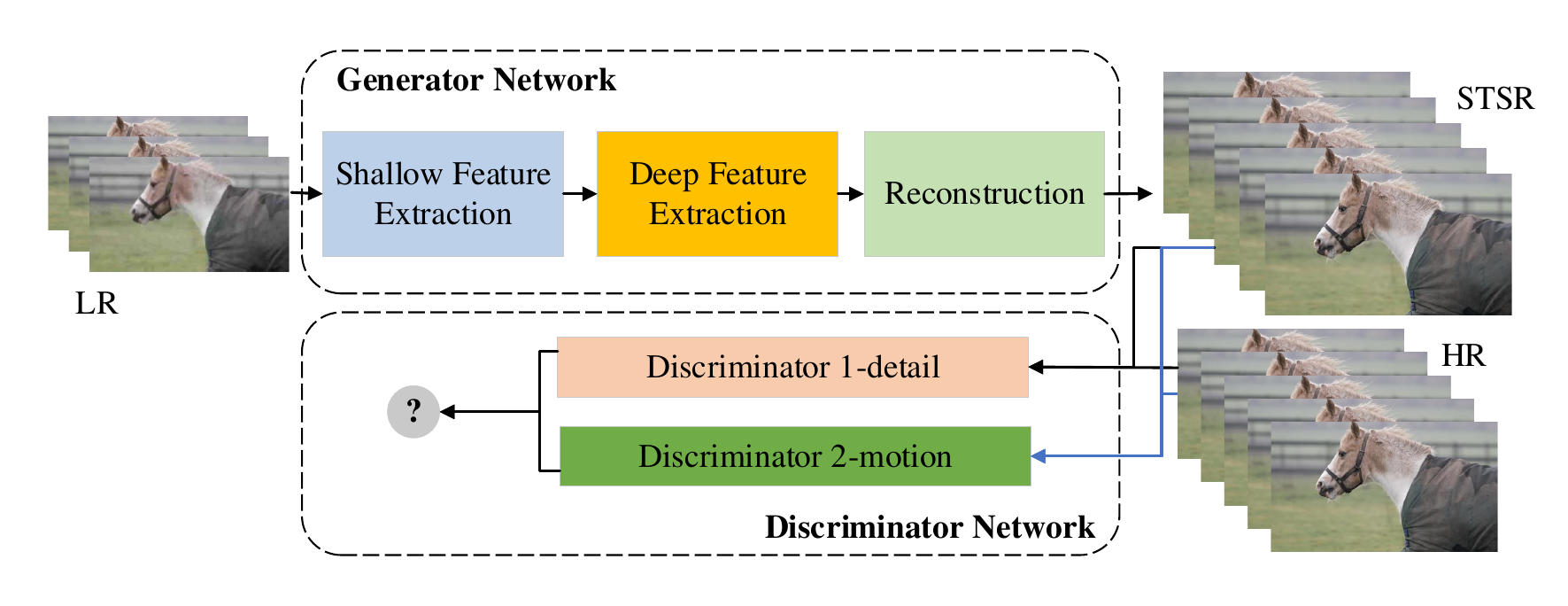}
	\caption{\textcolor{black}{Overview of the proposed 3DAttGAN. The network uses a generative adversarial network to transform a low-resolution, low-frame-rate input into a high-resolution, high-frame-rate output. The generative network consists of shallow feature extraction, deep feature extraction, and reconstruction. The discriminative network uses two criteria, one for the detail features and the other for the motion features, to accurately distinguish between STSR and HR.} }
	\label{overview}
\end{figure*}

The proposed 3D attention-based GAN (3DAttGAN) is shown in Fig.~\ref{overview}. The generative network consists of shallow feature extraction, deep feature extraction, and reconstruction. In the generative network, all the convolution operations are 3D convolutions. In the deep feature extraction module, a 3D attention mechanism is used to effectively capture temporal and spatial features. The discriminative network uses two criteria, one for the detail features and the other for the motion features, to accurately distinguish between STSR and HR.

\subsection{Proposed 3D Attention Mechanism}

Attention has a significant impact on human perception~\cite{28,29}. For example, humans can selectively focus on important visual features using partial glimpses for more efficient processing of visual structures~\cite{30}. 
The attention mechanism in a neural network is similar in essence to human visual attention, as it aims to select the most critical information for a given task from a vast amount of data.

\begin{figure}[htbp]  
	\centering
	\includegraphics [width=3in]{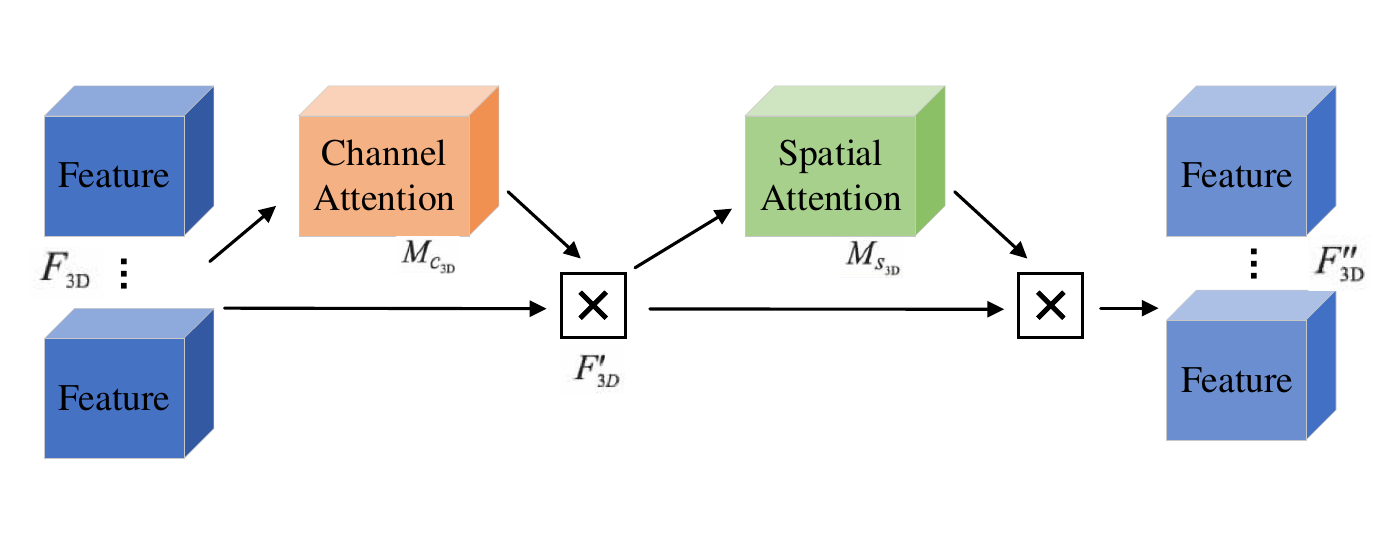}
	\caption{\textcolor{black}{Structure of 3D CSA. The module consists of channel attention and spatial attention. ``$\times$'' denotes element-wise multiplication. } }
	\label{3dcbam}
\end{figure}

The convolutional block attention module (CBAM)~\cite{CBAM} is a lightweight and effective module that can be directly applied to feed-forward convolutional neural networks. The module comprises two components: channel attention and spatial attention. 
CBAM generates attention maps on channel and spatial dimensions from the input feature map. It then uses these maps to adaptively refine the features via element-wise multiplication.

\textcolor{black}{
To fully exploit temporal and spatial information, we propose a 3D channel-spatial attention mechanism (3D CSA), as shown in Fig.~\ref{3dcbam}. 
Traditional attention mechanisms are designed for 2D convolutional neural networks and cannot be directly applied to 3D convolutions due to inconsistent dimensions. 3D convolutional networks have a depth dimension, and changes in depth should be considered when extracting spatial and depth features. 
}

\textcolor{black}{
For the features of an intermediate 3D convolution layer $\textit{\textbf{F}}_{3 D} \in \mathbf{R}^{W \times H \times D \times C}$, 3D CSA deduces the channel attention feature map $\textit{\textbf{M}}_{c_{3 \mathrm{D}}} \in \mathbf{R}^{1 \times 1 \times 1 \times C}$ and the spatial attention feature map $\textit{\textbf{M}}_{s_{3 \mathrm{D}}} \in \mathbf{R}^{1 \times H \times W \times C}$  in sequence. 
}

\begin{figure}[htbp]
	\centering
	\includegraphics [width=3in]{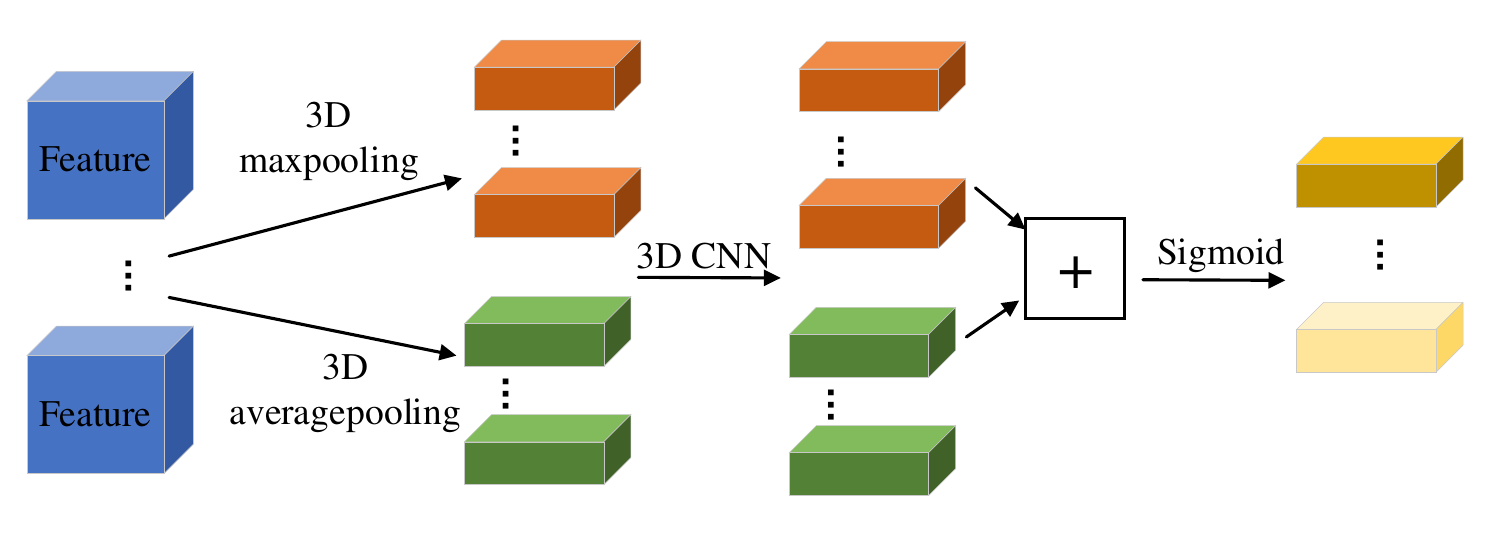}
	\caption{Channel attention module of 3D CSA. ``$ + $" adds the corresponding positions. Average pooling provides feedback for every pixel in the feature map, whereas max pooling only provides gradient feedback at the location with the maximum response in the feature map during gradient backpropagation.}
	\label{CA}
\end{figure}

The 3D CSA channel attention module identifies the channels that play a significant role in the final results in the fused 3D network, i.e., it selects the key features for prediction. The process is illustrated in Fig.~\ref{CA}. 
\textcolor{black}{
}

First, the input features $\textit{\textbf{F}}_{3 D}$ are separately processed by maximum pooling and average pooling based on width $W$, height $H$ and depth $D$. The 3D convolutional features are computed by element-wise addition. 
Next, a sigmoid activation is applied. By multiplying the generated channel feature map $\textit{\textbf{M}}_{c_{3 \mathrm{D}}}$ with the input feature $\textit{\textbf{F}}_{3 D}$, the final channel feature $\textit{\textbf{F}}_{3 \mathrm{D}}^{\prime}$ is generated as

\begin{equation}
	\textit{\textbf{M}}_{c_{3 \mathrm{D}}}\left(\textit{\textbf{F}}_{3 \mathrm{D}}\right) =\sigma\left(\operatorname{3DConv}\left(\operatorname{G}\left(\textit{\textbf{F}}_{3 \mathrm{D}}\right)\right)+\operatorname{3DConv}\left(\operatorname{P}\left(\textit{\textbf{F}}_{3 \mathrm{D}}\right)\right)\right.,\\
\end{equation}
where $ G(\cdot)$ and $P(\cdot) $ are 3D mean pooling and 3D maximum pooling operations, respectively, and $  \sigma$ is a sigmoid operation.

\begin{figure}[htbp]
	\centering
	\includegraphics [width=3in]{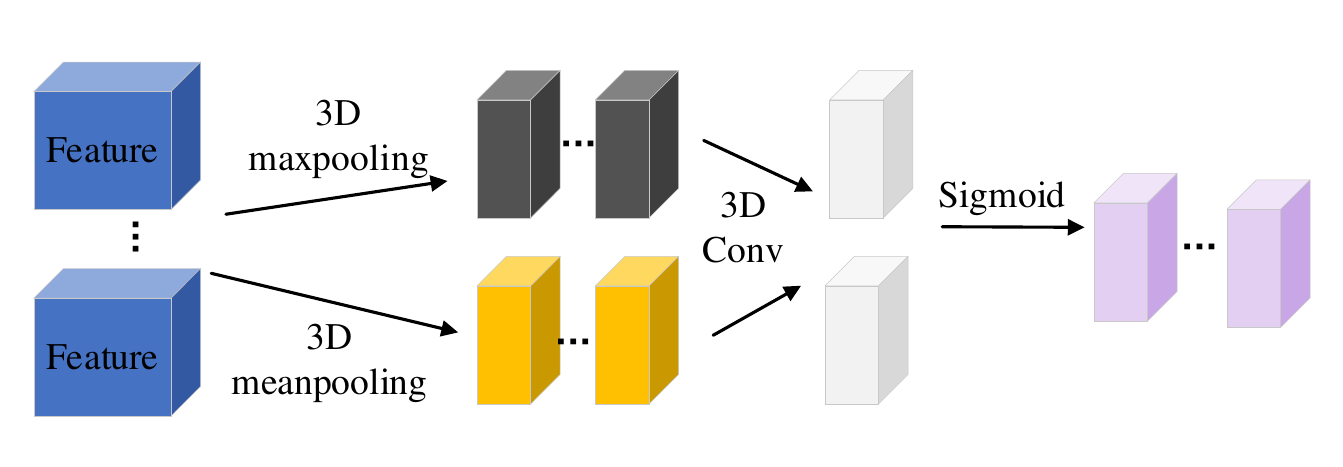}
	\caption{Spatial attention module of 3D CSA.}
	\label{SA}
\end{figure}

The 3D spatial attention model of 3D CSA focuses on the pixels in the image that have a significant impact in prediction. The attention feature extraction process is shown in Fig.~\ref{SA}.  
The spatial attention module takes the feature $ \textit{\textbf{F}}_{3 \mathrm{D}}^{\prime} $ of the channel attention module as input and applies channel-based 3D maximum pooling and 3D average pooling operations.
Next, the two extracted features are merged. 
Then, the dimension is reduced by the convolution operation, and the spatial attention feature map is generated through a sigmoid function. That is,

\begin{equation}
	\textit{\textbf{M}}_{s_{3 \mathrm{D}}}\left(\textit{\textbf{F}}_{3 \mathrm{D}}^{\prime}\right) =\sigma\left(f\left(\left[\operatorname{G}\left(\textit{\textbf{F}}_{3 \mathrm{D}}^{\prime}\right) ; \operatorname{P} \left(\textit{\textbf{F}}_{3 \mathrm{D}}^{\prime}\right)\right]\right)\right), \\
\end{equation}
where $ f $ denotes the convolution operation. 

Finally, the output feature map and the input $ \textit{\textbf{F}}_{3 \mathrm{D}}^{\prime} $ are multiplied pointwise to generate the final feature $ \textit{\textbf{F}}_{3 \mathrm{D}}^{\prime \prime} $. The overall process can be expressed as
\begin{equation}
	\begin{aligned}
		&\textit{\textbf{F}}_{3 \mathrm{D}}^{\prime}=\textit{\textbf{M}}_{c_{3 \mathrm{D}}}\left(\textit{\textbf{F}}_{3 \mathrm{D}}\right) \otimes \textit{\textbf{F}}_{3 \mathrm{D}} , \\
		&\textit{\textbf{F}}_{3 \mathrm{D}}^{\prime \prime}=\textit{\textbf{M}}_{s_{3 \mathrm{D}}}\left(\textit{\textbf{F}}_{3 \mathrm{D}}^{\prime}\right) \otimes \textit{\textbf{F}}_{3 \mathrm{D}}^{\prime},
	\end{aligned}
\end{equation}
where $  \otimes $ represents pointwise multiplication.

\subsection{Proposed Generative Adversarial Network}

Generative adversarial networks (GANs)~\cite{8} are highly efficient models that can learn complex probability distributions from a given prior distribution. GANs were initially developed for the purpose of generating images~\cite{8}. Since then, GANs have been used for many generative tasks, including audio synthesis, 3D modeling, and image-to-image translation tasks. 
The exceptional generative capability of the GANs has been used to generate high-quality images for various reconstruction tasks, e.g.,~\cite{9,10,11}.

\textcolor{black}{
GANs and Variational Autoencoders (VAEs) are two commonly used generative models in image and video processing. GANs are better suited for tasks that demand extremely high-quality details, whereas VAEs are more suitable for tasks that emphasize stability and interpretability. GANs operate by training a generator and a discriminator network to produce high-quality images or videos. They excel in super-resolution tasks, yielding sharper images by capturing fine textures and structures. On the other hand, VAEs may produce blurry results in super-resolution tasks, as they tend to generate smoother images with less focus on details. VAEs typically impose higher quality requirements on input images or videos, potentially necessitating more training data and complex network architectures to achieve high-quality super-resolution results. Our objective is to obtain higher quality videos, including fine details and textures. Therefore, we have opted for GANs instead of VAEs.
} 

\begin{figure}[htbp]
	\centering
	\includegraphics [width=2.5in]{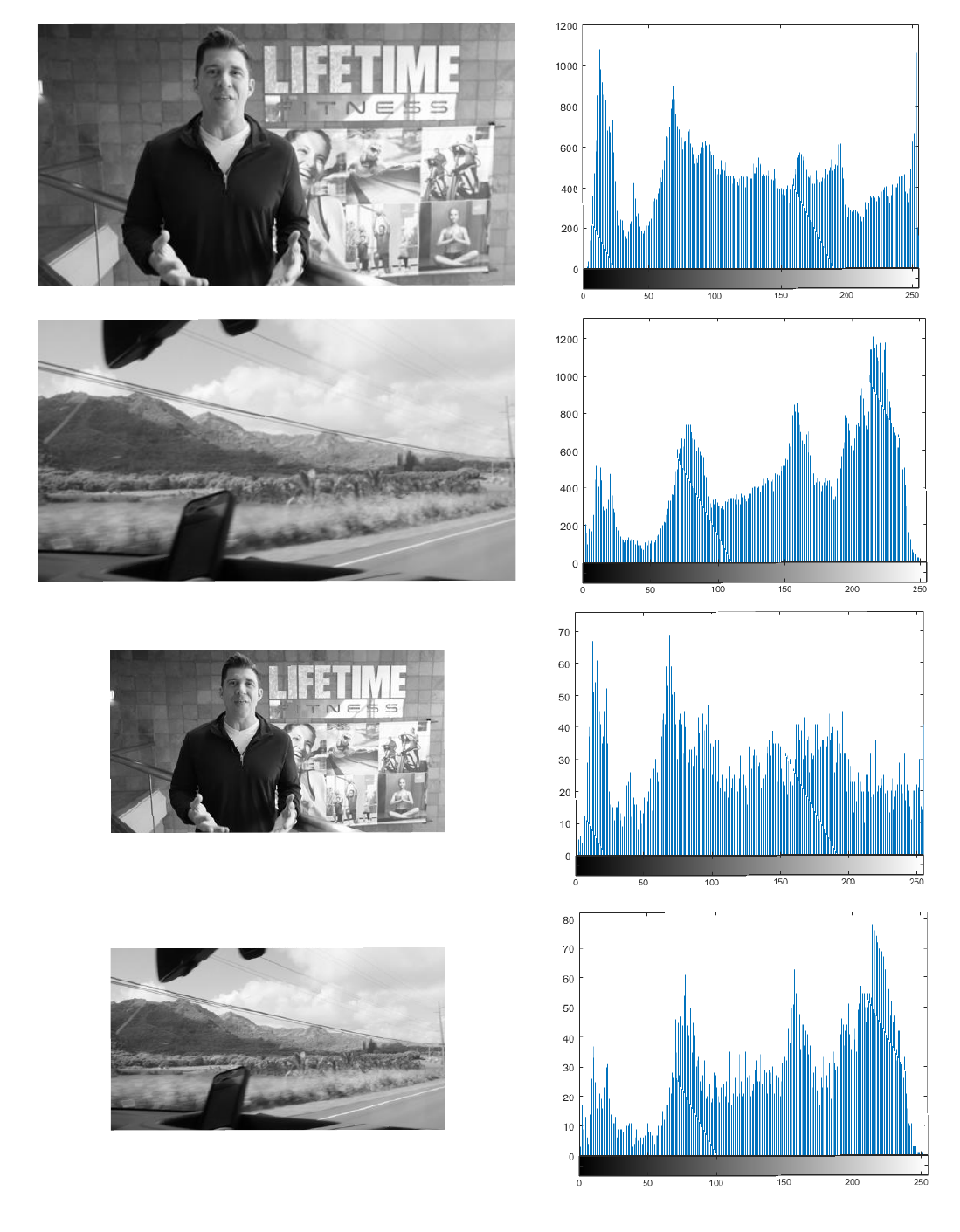}
	\caption{Histograms of different images. In the first column, the first two images are the original images, while the last two images are their subsampled counterparts. The second column shows the corresponding histograms.}
	\label{image-hist}
\end{figure}

Our goal is to improve the resolution of LR video frames. Fig.~\ref{image-hist} shows the histogram distribution of various images. The distribution of the same image is roughly the same across different resolutions, while the distribution between different images varies widely. 
\textcolor{black}{
When we super-resolve a video, the SR video and the original video are in similar domains, which facilitates convergence of the neural network. 
In the proposed STSR task, GANs can automatically learn the data distribution through generative adversarial training, enabling the generation of content that closely matches the input data.}

\subsubsection{Generative Network}

\begin{figure*}[htbp]
	\centering
	\includegraphics [scale=0.5]{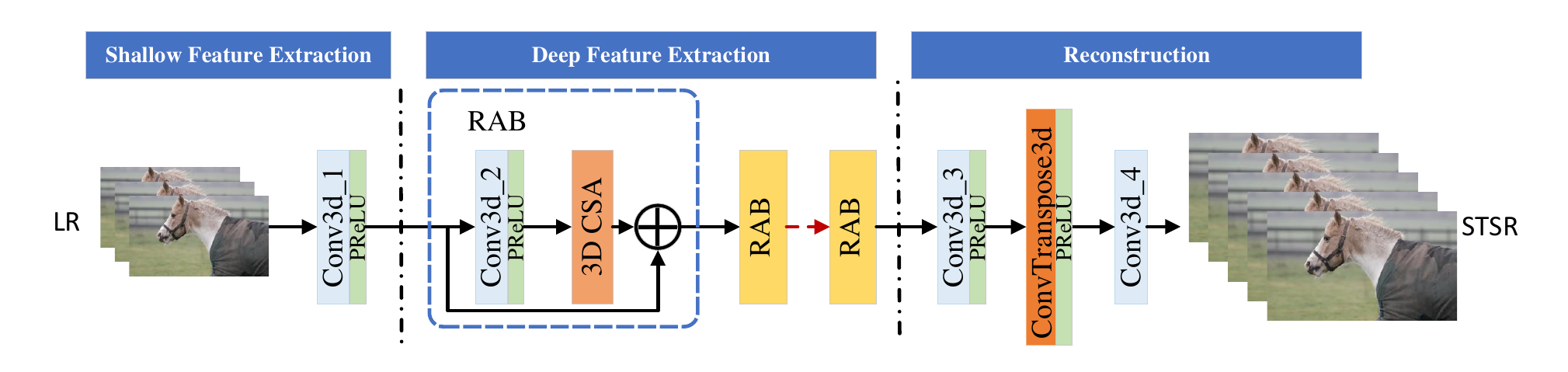}
	\caption{\textcolor{black}{Structure of the generator. The generative network consists of shallow feature extraction, deep feature extraction, and reconstruction with 3D convolutions. Each RAB consists of a 3D convolution (Conv3d), a parametric rectified linear unit (PReLU), and a 3D attention module (3D CSA). ConvTranspose3d upsamples the features to the required resolution and activates them.}}
	\label{GN}
\end{figure*}

The generative network is the core of our video STSR method. We use a 3D convolution to achieve STSR and a 3D attention mechanism to further improve the results. The network architecture is shown in Fig.~\ref{GN}. 

\textcolor{black}{
Shallow feature extraction uses a single-layer 3D convolution to extract features from the input video frame. This results in a small overlap area in the receptive field, which enables the network to extract fine details. Shallow feature extraction is followed by deep feature extraction through multiple residual attention convolution blocks (RABs).
Each RAB consists of a 3D convolution, an activation function, and a 3D CSA module. }
As the input contains both spatial and temporal information, a 3D convolution is used to effectively extract the corresponding features simultaneously.
The activation function uses a parametric rectified linear unit(PReLU)~\cite{42} which can improve the fitting ability of the model and mitigate the overfitting risk without extra parameters. 
3D CSA dynamically selects important information from the input features by adapting the weight of each feature. 

The deep feature extraction module is composed of $N$ RABs, as shown in Fig.~\ref{GN}. The output of the $n$-th RAB $\textit{\textbf{F}}_{n}$ is
\begin{equation}
	\begin{aligned}
		\textit{\textbf{F}}_{n} &=R_{\text {RAB}}\left(\textbf{\textit{F}}_{n-1}\right) \\
		&=C_{\text {3DCSA}}\left(\operatorname{3DConv}\left(\textit{\textbf{F}}_{n-1}\right)\right)+\textit{\textbf{F}}_{n-1},
	\end{aligned}
	\label{eq}
\end{equation}
where $n \in\{1, \cdots, N\}$, $R_{\text {RAB}}$ denotes the feature extraction operator of the RAB, ${\textbf{\textit{F}}_{n-1}}$ is the input feature of the RAB,  $\operatorname{3DConv}(\cdot)$ denotes the 3D convolution followed by a PReLU activation operation, and $C_{\text {3DCSA}}(\cdot)$ is the 3D CSA operator.

\begin{figure}[htbp]
	\centering
	\includegraphics [width= 3.2in]{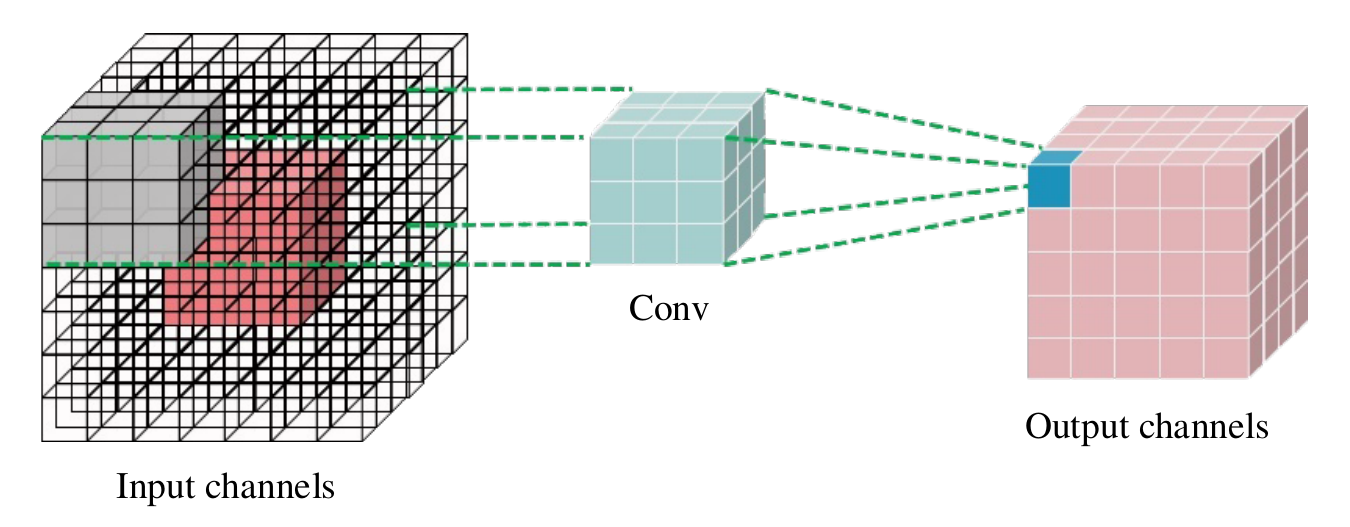}
	\caption{ConvTranspose3d schematic.}
	\label{ConvT3d}
\end{figure}

\textcolor{black}{
The reconstruction module is mainly composed of ConvTranspose3d and Conv3d. ConvTranspose3d upsamples the features to the required resolution and activates them, while Conv3d reconstructs the video frames. ConvTranspose3d is illustrated in Fig.~\ref{ConvT3d}. When both space and time resolutions are increased, STSR is achieved. When only the former is increased, SSR is achieved. When only the latter is increased, TSR is achieved.
}

\subsubsection{Discriminative Network}

\begin{figure*}[htbp]
	\centering
	\includegraphics [scale=0.6]{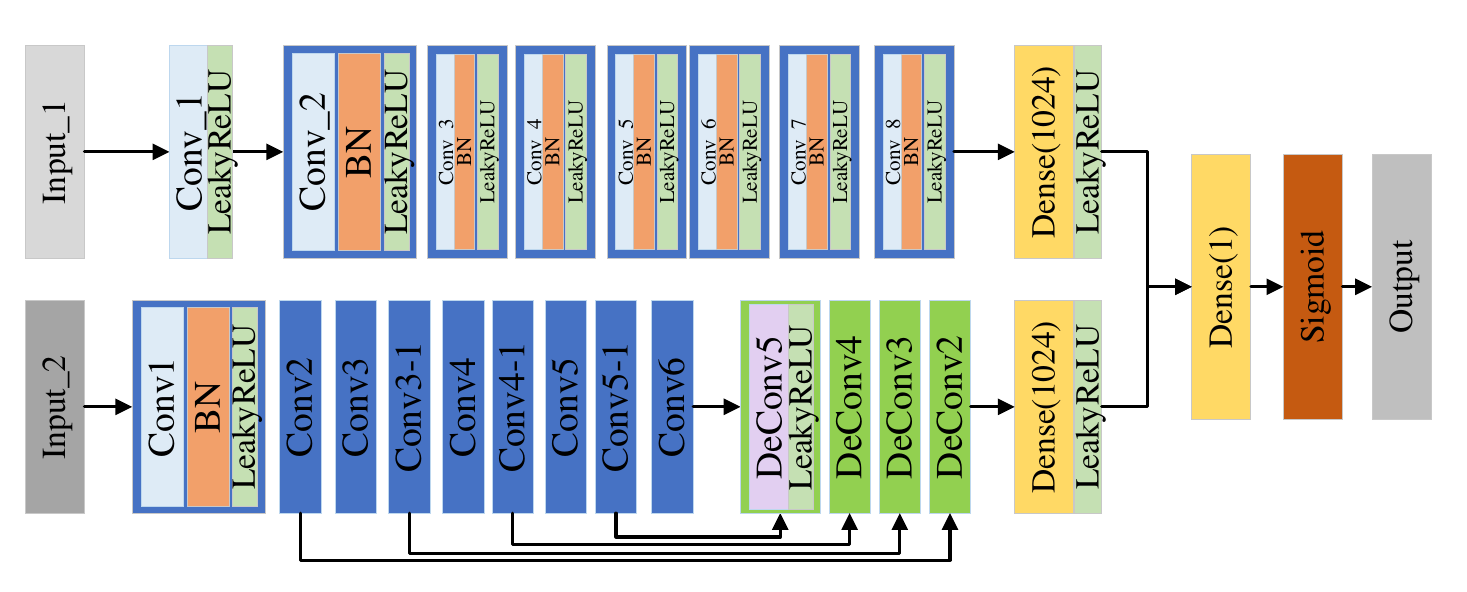}
	\caption{\textcolor{black}{Structure of the discriminator. Input\_1 is a single video frame. Input\_2 are two successive video frames. The blue boxes on the top branch have three layers: a 2D convolution layer, a batch normalization (BN) layer, and a LeakyReLU activation layer.}}
	\label{DN}
\end{figure*}

The proposed discriminator (Fig.~\ref{DN}) aims to accurately distinguish between SR and true HR frames.
There are two discriminant criteria: the characteristics of the video frame itself and the motion information between frames. 
The discriminator is composed of two branches. The top branch is used for detailed texture judgment, while the bottom branch compares motion information. 
The top branch extracts detailed information from each video frame, while the bottom branch processes two successive video frames to obtain their optical flow motion information. As the two branches of the discriminator deal with the texture detail and motion information separately, a 2D convolution is used for all the convolution layers of the discriminator. The activation function used is a leaky
rectified linear unit (LeakyReLU) layer.  
A full connection layer is used to integrate the outputs of the two branches.
Finally, a sigmoid function is used to judge whether the input video is a real HR video or not.

\subsubsection{Loss function}

The loss function is the sum of two loss functions: $ L_{S R} $ for SR loss and $ L_{G A N} $ for the GAN discriminator loss. $ L_{SR} $ calculates the L2 distance between the prediction result $ I_t^{S R} $ and the ground truth $ I_t^{H R} $:
\begin{equation}
	L_{S R}=\left\|I_t^{S R}-I_t^{H R}\right\|_2^2.
\end{equation}

Inspired by least squares GAN (LSGAN)~\cite{47}, we use the least-squares loss instead of the cross-entropy loss as the loss function for the discriminator. 
\textcolor{black}{
The least-squares discriminator loss incentivizes the discriminator to produce continuous outputs that closely align with the true distribution.
This fosters the production of smoother discriminator outputs, enhancing the discriminator's ability to assess the quality of the generated samples more effectively.
Conventional GANs may face the issue of mode collapse during training.
The utilization of the least-squares loss mitigates this problem, rendering the generator more inclined to generate diverse samples.
Simultaneously, the least-squares loss typically exhibits greater stability and is less susceptible to the vanishing gradient problem throughout the training process.
Consequently, this facilitates the training of GANs and generally results in higher-quality generated samples.
}

The discriminator loss function is thus~\cite{47}

\begin{equation}
	\begin{aligned}
		L_{G A N} &=\mathbb{E}_{I_t^{H R} \sim P_{\left(I_t^{H R}\right)}}\left(D\left(I_t^{H R}\right)-b\right)^2 \\
		&+\mathbb{E}_{z \sim P_z}\left(D\left(I_t^{S R}\right)-a\right)^2,
	\end{aligned}
\end{equation}
where $ D $ is the discriminator, the generated sample and the real sample are coded as $ a=0, b=1 $ respectively, $ z $ represents random noise that follows a Gaussian distribution $ P_z $, and $ P_{\left(I_t^{H R}\right)} $ is the distribution of $ I_t^{H R} $.

\section{Experimental Results}
\begin{table*}[!b]
	\centering
	\caption{STSR quality on Vimeo90K.}
	\renewcommand\arraystretch{1.5}  
		\begin{tabular}{l|c|c||l|c|c}
			\hline
			\makecell{Method \\ (SSR+TSR/STSR)} & PSNR↑ & SSIM↑ & 
			\makecell{Method \\(TSR+SSR/STSR)} & PSNR↑ & SSIM↑  \\
			\hline
			DBPN~\cite{48}+ToFlow~\cite{34} & 29.87 & 0.905 & ToFlow~\cite{34}+DBPN~\cite{48} & 28.82 & 0.887 \\
			TDAN~\cite{13}+ToFlow~\cite{34} & 30.41 & 0.918 & ToFlow~\cite{34}+TDAN~\cite{13} & 29.65 & 0.908 \\
			BasicVSR~\cite{15}+ToFlow~\cite{34} & 30.58 & 0.924 & ToFlow~\cite{34}+BasicVSR~\cite{15} & 30.22 & 0.915 \\
			LatticeNet~\cite{LatticeNet}+ToFlow~\cite{34} & 30.84 & 0.928 & ToFlow~\cite{34}+LatticeNet~\cite{LatticeNet} & 30.62 & 0.924 \\
			SwinIR~\cite{SwinIR}+ToFlow~\cite{34} & 30.91 & 0.929 & ToFlow~\cite{34}+SwinIR~\cite{SwinIR} & 30.73 & 0.926 \\
			DBPN~\cite{48}+DAIN~\cite{18} & 30.25 & 0.916 & DAIN~\cite{18}+DBPN~\cite{48} & 29.42 & 0.906 \\
			BasicVSR~\cite{15}+DAIN~\cite{18} & 30.72 & 0.925 & DAIN~\cite{18}+BasicVSR~\cite{15} & 30.45 & 0.919 \\
			LatticeNet~\cite{LatticeNet}+CAIN~\cite{CAIN} & 30.95 & 0.929 & CAIN~\cite{CAIN}+LatticeNet~\cite{LatticeNet} & 30.82 & 0.922 \\
			SwinIR~\cite{SwinIR}+EDSC~\cite{EDSC} & 30.98 & 0.930 & EDSC~\cite{EDSC}+SwinIR~\cite{SwinIR} & 30.84 & 0.928 \\
			\hline
			STVUN~\cite{b1} &   29.68  &  0.908 & STARnet~\cite{STSR} &   30.81    &  0.924 \\
			TMNet~\cite{b5} &   30.92  &   0.928  & Ours  &  \textbf{31.86}     & \textbf{0.945} \\
			\hline
		\end{tabular}%
	\label{ST-SR}%
\end{table*}%

\subsection{Data sets and training configuration}

The proposed network was trained using the training-subset of Vimeo-90K dataset~\cite{43} which includes more than 60000 videos of size 7 (number of frames) $ \times $ 448 (spatial width) $ \times $ 256 (spatial height). The Vid4 dataset~\cite{44} and the Vimeo-90K~\cite{43} testing dataset were used for testing. For training, we first generated LR videos by spatial and temporal subsampling. Specifically, the spatial resolution was down-sampled to 112$ \times $64 through bi-cubic interpolation, while the even frames were deleted. 
To evaluate the quality of the generated full-resolution videos, we used the peak signal-to-noise ratio (PSNR) and the structural similarity index (SSIM)~\cite{46}.
As the proposed STSR method can generate high spatial-resolution (SSR) and high frame-rate (TSR) videos simultaneously, the SSR and TSR results are also presented.

The proposed model was developed on the PyTorch platform and trained with a 2080Ti GPU. In the implementation, we randomly cropped the videos for training to patches of size 32$ \times $32$ \times $4 and used ground truth video patches of size (32$ \times $4)$ \times $(32$ \times $4)$ \times $7 as the labels.

\subsection{Comparison with state-of-the-art methods}

As benchmarks for STSR methods, we used the state-of-the-art (SOTA) STSR methods STARnet~\cite{STSR}, STVUN~\cite{b1}, and TMNet~\cite{b5}, as well as a combination of SOTA SSR and TSR methods. 
\textcolor{black}{
If the source code of the SOTA method was available, we used it; otherwise, we implemented the method ourselves. In this way, all comparison methods were trained and tested in the same way as our method. 
For these combinations, we used DBPN~\cite{48}, 
TDAN~\cite{13}, BasicVSR~\cite{15}, LatticeNet~\cite{LatticeNet} and SwinIR~\cite{SwinIR} for SSR, and ToFlow~\cite{34}, DAIN~\cite{18}, CAIN~\cite{CAIN} and EDSC~\cite{EDSC} for TSR.
Table~\ref{ST-SR} shows the results for Vimeo-90K. We can see that the combination of SSR+TSR performs better than TSR+SSR. We can also see that the proposed method had both the highest PSNR (31.68 dB on average) and SSIM (0.945).
}

\textcolor{black}{
To further verify the effectiveness of proposed method, we tested it on the REDS dataset~\cite{REDS}. The experimental results are shown in Table~\ref{REDS}.
As can be seen from the table, the proposed method consistently maintained the best performance.
To evaluate the performance under at motion conditions, we divided, as in~\cite{34}, the Vimeo-90K dataset into three groups based on the level of motion: fast, medium, and slow. These groups comprise 1225, 4977, and 1613 video clips, respectively. The gains were largest in the fast motion group (0.94 dB on average), see Table~\ref{Com}.
}

\begin{table}[htbp]
	\centering
	\caption{Comparison of STSR results on REDS. }
	\renewcommand\arraystretch{1.5}  
	\begin{tabular}{c|c|c||c|c|c}
		\hline
		Method  & PSNR & SSIM & 
	    Method  & PSNR & SSIM  \\
	    \hline
	STVUN~\cite{b1} &   27.35  &  0.872 & STARnet~\cite{STSR} &   28.48    &  0.881 \\
	TMNet~\cite{b5} &   28.73  &   0.887  & Ours  &  \textbf{29.21}     & \textbf{0.893} \\
	\hline
	\end{tabular}
	\label{REDS}
\end{table}

\begin{table}[htbp]
	\centering
	\caption{Comparison of STSR results on classified Vimeo90K. }
	\renewcommand\arraystretch{1.5}  
	\begin{tabular}{c|c|c|c|c|c|c}
		\hline
		\multirow{2}{*}{Method} & \multicolumn{2}{c|}{ Fast motion }  & \multicolumn{2}{c|}{ Medium motion }& \multicolumn{2}{c}{ Slow motion }\\
		\cline{2-7}
		& PSNR & SSIM & PSNR & SSIM & PSNR & SSIM \\
		\hline
		STARnet~\cite{STSR} & 30.63 & 0.924 & 30.74 & 0.928 & 30.86 & 0.931\\
		TMNet~\cite{b5} & 30.89 & 0.930 & 30.97 & 0.938 & 31.04 & 0.940\\
		Ours   & \textbf{31.83} & \textbf{0.943} & \textbf{31.85} & \textbf{0.945} & \textbf{31.88} & \textbf{0.948} \\
		\hline		
	\end{tabular}
	\label{Com}
\end{table}
\begin{table*}[!b]
	\centering
	\caption{SSR quality on Vimeo90K.}
	\renewcommand\arraystretch{1.5}  
		\begin{tabular}{c|c|c|c|c|c|c|c|c}
			\hline
			Method &  Bicubic & DBPN~\cite{48} &  RBPN~\cite{49} & TDAN~\cite{13} & BasicVSR~\cite{15} & SwinIR~\cite{SwinIR} & Ours & Ours-SSR \\ 
			\hline
			PSNR      & 27.89 & 29.83  & 30.97 & 32.42 & 33.02 &  33.16 &  32.55 & \textbf{33.23}\\
			SSIM      & 0.887 & 0.913  & 0.938 & 0.943 & 0.957 &  0.959 & 0.942 & \textbf{0.961}\\
			\hline
		\end{tabular}
	\label{SSR}
\end{table*}
\begin{table*}[!b]
	\centering
	\caption{TSR quality for the original resolution on Vimeo90K.}
	\renewcommand\arraystretch{1.5}  
		\begin{tabular}{c|c|c|c|c|c|c}
			\hline
			Method &  DVF~\cite{16} & ToFlow~\cite{34} & MEMC-Net~\cite{17} & DAIN~\cite{18} & EDSC & Ours-TSR  \\
			\hline
			PSNR      & 31.54 & 33.73 & 34.29 & 34.71 & 34.83 & \textbf{34.93}  \\
			SSIM      & 0.946 & 0.968 & 0.974 & 0.976 & 0.976 & \textbf{0.978}  \\
			\hline
		\end{tabular}
	\label{TSR}
\end{table*}

\begin{table*}[!b]
	\centering
	\caption{Computational complexity.}
	\renewcommand\arraystretch{1.5}  
	\resizebox{\linewidth}{!}{
		\begin{tabular}{c|c|c|c|c|c|c|c|c|c}
			\hline
			Methods & {ToFlow-DBPN}& {DBPN-ToFlow} & {DBPN-DAIN} & {DBPN\_{MI}-DAIN} & {DAIN-RBPN} & {RBPN-DAIN} & {TDAN-DAIN} & {STARnet} & Ours \\
			\hline
			Parameters     & 20.2M & 20.2M & 38.4M & 40.2M & 36.7M & 36.7M & 26.2M & \textbf{19.2M}  & 20.3M \\
			Runtime(s)      & 3.44 & 3.85 & 3.26 & 3.42 & 3.71& 4.26 & 3.52 & 3.25 & \textbf{3.14}\\
			\hline
		\end{tabular}
	}
	\label{computational}
\end{table*}

\textcolor{black}{
	Our method allows us to achieve SSR in two ways. The first one (named Ours in Table~\ref{SSR}) is to consider SSR as a byproduct of STSR. The second one (named Ours-SSR in Table~\ref{SSR}) is to convert our STSR method into an SSR one by keeping the same temporal resolution and modifying only the spatial resolution in Convtranspose3d. Table~\ref{SSR} compares these two approaches to five SOTA methods (DBPN~\cite{48}, RBPN~\cite{49}, TDAN~\cite{13}, BasicVSR~\cite{15}), and SwinIR~\cite{SwinIR}. The results show that the first approach did not give competitive results. However, the second approach achieved better results than the current SOTA SSR methods. 
}

\textcolor{black}{
	Similarly, keeping the same spatial resolution and modifying only the temporal resolution in Convtranspose3d (named Ours-TSR in Table~\ref{TSR}) allowed us to outperform five TSR methods (DVF~\cite{16}, ToFlow~\cite{34}, MEMC Net~\cite{17}, DAIN~\cite{18}, and EDSC~\cite{EDSC}), see Table~\ref{TSR}. 
}

To synthesize high-quality video frames, SSR and TSR networks often require very complex reconstruction blocks. Therefore, the two-stage STSR networks have a huge number of parameters, leading to a high computational cost. Table~\ref{computational} compares the number of network parameters and processing time of various methods. 
\textcolor{black}{
	Since our method consists of two networks (a generator and a discriminator), it includes more parameters than methods like STARnet that consist of a single network. However, because the generator is simpler than these networks, and our method only uses the generator in the testing phase, its running time is shorter.}

\begin{figure*}[htbp]
	\centering
	\includegraphics [scale=1.5]{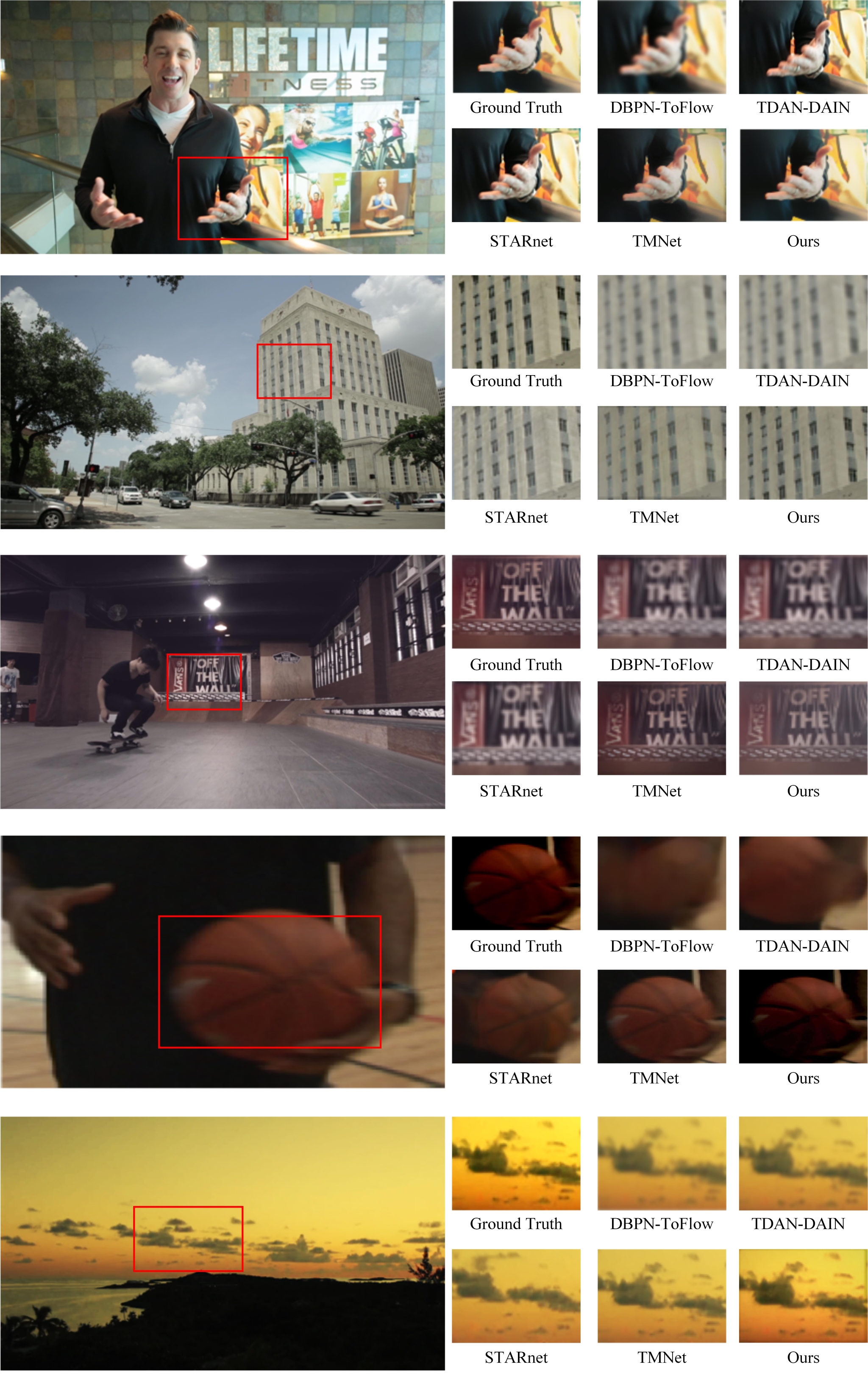}
	\caption{Visual results. The red squares indicate the highlighted areas.}
	\label{Vis}
\end{figure*}

Finally, Fig.~\ref{Vis} compares the visual results. The proposed method can provide better visual quality, especially for videos with rich textures and fast motion.

\subsection{Ablation study}

\subsubsection{Effectivness of 3D CSA}

\begin{figure}[htbp]
	\centering
	\includegraphics [width= 3in]{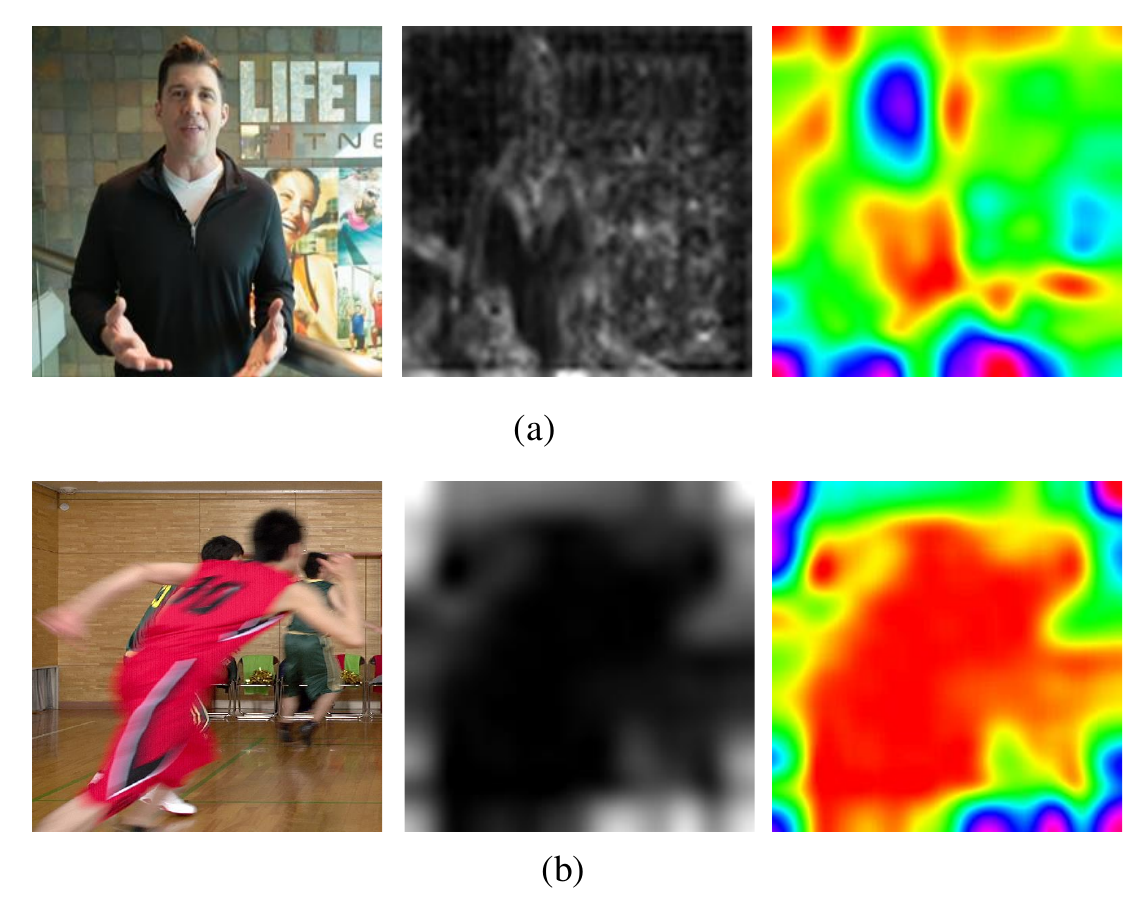}
	\caption{Visualization of network features after 3D CSA. (a) shows a frame of a low motion video, (b) shows a frame of a high motion video. The image shown on the left is the original image, the gradient-weighted class activation map is in the middle, and the heat map is on the right.  }
	\label{Vis3d}
\end{figure}

Attention mechanisms not only guide the model's focus, but also enhance the representation of relevant information.
The aim is to enhance representation capacity with the help of an attention mechanism that prioritizes salient features and disregards irrelevant ones.
Fig.~\ref{Vis3d} shows the feature map and the heat map generated by 3D CSA. 
Fig.~\ref{Vis3d}(a) shows that in a low motion sequence, feature extraction is mainly focused on areas with complex textures, such as faces and contours, while less attention is paid to the background area. The results show that 3D CSA can effectively extract important information from spatial features. Fig.~\ref{Vis3d}(b) corresponds to a video frame with large motion. We can clearly see that the large motion is prominently represented. The result shows that 3D CSA can also effectively capture motion information in video frames.

\begin{figure*}[htbp]
	\centering
	\includegraphics [scale=0.65]{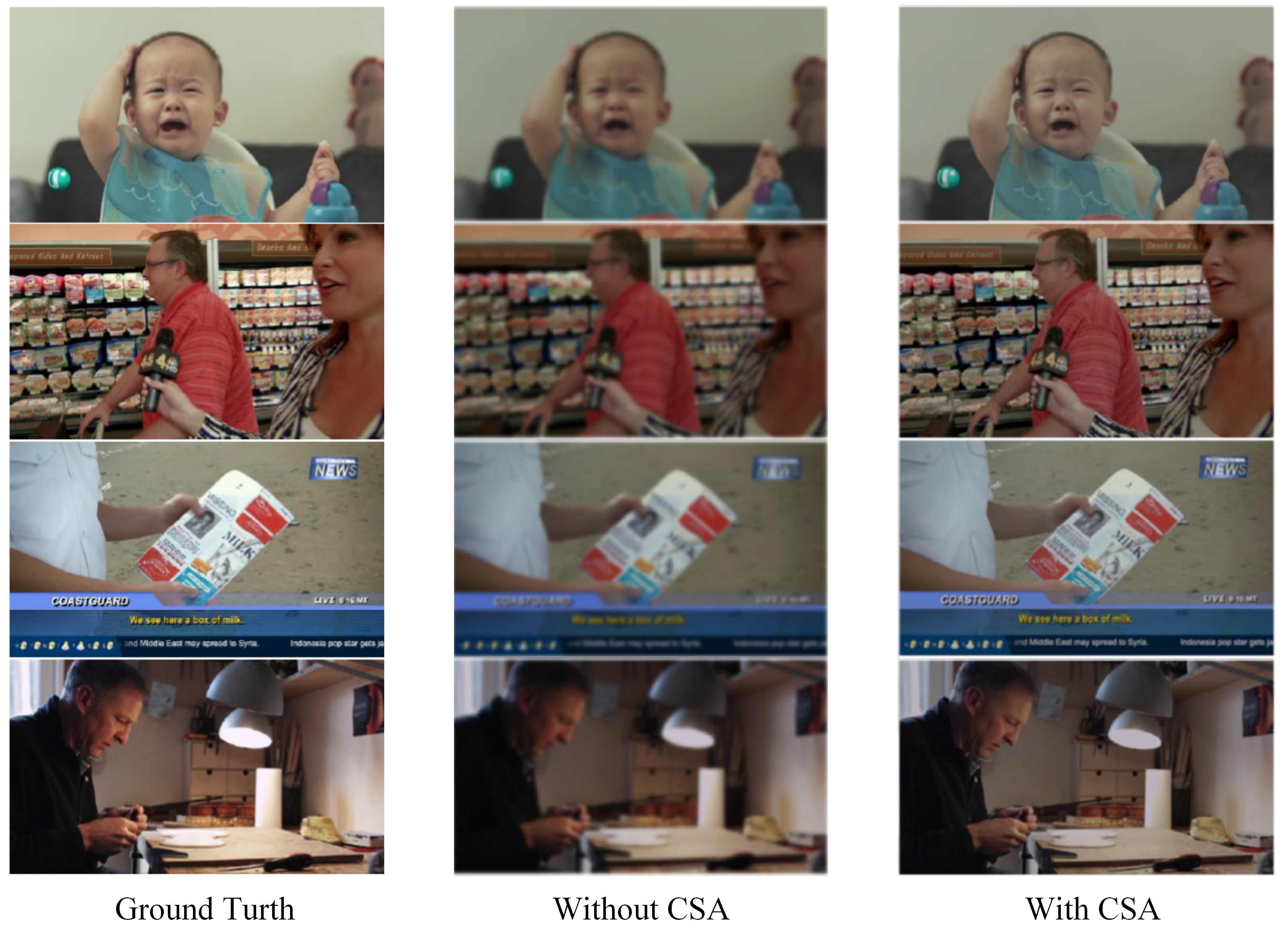}
	\caption{\textcolor{black}{Visual results with and without 3D CSA. The left column shows the ground truth, the middle column shows the result when 3D CSA is not used, and the right column shows the result when 3D CSA is used.}}
	\label{w/o 3D CSA}
\end{figure*}

\begin{table}[htbp]
	\centering
	\caption{Effectiveness of 3D CSA on Vimeo90K. The first row shows the results without 3D CSA, while the second row shows the results with 3D CSA.}
	\renewcommand\arraystretch{1.5}  
	\resizebox{\linewidth}{!}{ 
		\begin{tabular}{c|c|c|c|c|c|c}
			\hline
			\multirow{2}{*}{Method} & \multicolumn{2}{c|}{  STSR }  & \multicolumn{2}{c|}{ SSR }& \multicolumn{2}{c}{ TSR }\\
			\cline{2-7}
			& PSNR & SSIM & PSNR & SSIM & PSNR & SSIM \\
			\hline
			w/o       & 30.21 & 0.8342 & 31.15 & 0.8794 & 28.95 & 0.8197\\
			with     & \textbf{30.90} & \textbf{0.8985} & \textbf{31.80} & \textbf{0.9222} & \textbf{29.62} & \textbf{0.8750}   \\
			\hline
		\end{tabular}
	}
	\label{3D CSA}
\end{table}

Table~\ref{3D CSA} illustrates the importance of 3D CSA. We can see that the average PSNR increased by 0.67 dB when using 3D CSA. Fig.~\ref{w/o 3D CSA} shows a visual quality comparison between the models with and without 3D CSA. We can see that the image quality can be significantly enhanced with 3D CSA.

\subsubsection{Effect of the number of RABs }

The proposed RAB, which includes 3D CSA, is the most important part of the proposed generator network. To assess the impact of RAB, we experimentally studied the influence of the number of RABs, as shown in Tables~\ref{numBAB-90} and~\ref{numBAB-4}. In general, the quality improved with more RABs. For the test set Vimeo90K, when the number of RABs was increased from 3 to 7, the improvement was 0.51 dB for STSR, 0.95 dB for SSR, and 0.19 dB for TSR. When the number of RABs was raised from 9 to 12, the improvements were 0.46 dB for STSR, 0.56 dB for SSR, and 0.39 dB for TSR. For the Vid4 test set, the overall performance did not improve beyond a certain number of RABs. Specifically, when the number of RABs was raised from 9 to 12, the improvement was only 0.01 dB for STSR, 0.01 dB for SSR, and 0.02 dB for TSR.  
\textcolor{black}{This may be because the network reaches an overfitting state for Vid4 when the number of RABs reaches a certain level. Consequently, when the number of RABs further increases, the network's performance starts to stagnate.}
To balance computational cost and performance, we set the number of RABs to 9 for Vid4 and 12 for Vimeo90K.
\begin{table}[h]
	\centering
	\caption{Comparison of the effect of the number of RABs on Vimeo90K.}
	\renewcommand\arraystretch{1.5}
	\resizebox{\linewidth}{!}{ 
		\begin{tabular}{c|c|c|c|c|c|c}
			\hline
			\multirow{2}{*}{RABs} & \multicolumn{2}{c|}{  STSR }  & \multicolumn{2}{c|}{ SSR }& \multicolumn{2}{c}{ TSR }\\
			\cline{2-7}
			& PSNR & SSIM & PSNR & SSIM & PSNR & SSIM \\
			\hline
			3       & 29.80 & 0.8775 & 30.29 & 0.8863 & 28.85 & 0.8249\\
			7       & 30.31 & 0.8824 & 31.24 & 0.9065 & 29.04 & 0.8670   \\
			9       & 30.90 & 0.8985 & 31.80 & 0.9222 & 29.62 & 0.8750   \\
			12  & \textbf{31.36} & \textbf{0.9002} & \textbf{32.36} & \textbf{0.9324} & \textbf{30.01} & \textbf{0.8824} \\
			\hline
	\end{tabular}}
	\label{numBAB-90}
\end{table}
\begin{table}[h]
	\centering
	\caption{Comparison of the effect of the number of RABs on Vid4.}
	\renewcommand\arraystretch{1.5}  
	\resizebox{\linewidth}{!}{
		\begin{tabular}{c|c|c|c|c|c|c}
			\hline
			\multirow{2}{*}{RABs} & \multicolumn{2}{c|}{  STSR }  & \multicolumn{2}{c|}{ SSR }& \multicolumn{2}{c}{ TSR }\\
			\cline{2-7}
			& PSNR & SSIM & PSNR & SSIM & PSNR & SSIM \\
			\hline
			3       & 29.75 &         0.8734  & 30.16           & 0.8572 & 28.56 & 0.8175\\
			7       & 30.56 &         0.8804  & 31.02           & 0.8867 & 28.94 & 0.8256   \\
			9       & 30.71 & \textbf{0.8955} & 31.86 & 0.9013 & 29.48 & 0.8376   \\
			12      & \textbf{30.72} & 0.8953 & \textbf{31.87}  &\textbf{0.9014} & \textbf{29.50} & \textbf{0.8398} \\
			\hline
		\end{tabular}
	}
	\label{numBAB-4}
\end{table}

\subsubsection{Effectiveness of the discriminator} 

To evaluate the efficacy of our generative adversarial structure, we compared the results of the generator without the proposed discriminator (only generator), with only the top branch of the proposed discriminator (only discriminator 1), with only the bottom branch of the proposed discriminator (only discriminator 2), and with the proposed GAN structure (full GAN). The results of this experiment are given in Table~\ref{CGAN}. From this table, we can see that the full GAN structure improved the performance of the structure based on only the generator by 0.5 dB, 0.59 dB, and 0.37 dB for STSR, SSR and TSR, respectively, confirming the efficacy of the proposed design. Comparing the results in the middle two rows of Table~\ref{CGAN}, we see that better SSR results were obtained when only the top branch of the discriminator was used, and better TSR results were obtained when only the bottom branch of the discriminator was used. The results motivate the structure of the discriminator, with a top branch for texture features and a bottom one for motion information.

\begin{table}[htbp]
	\centering
	\caption{Comparison results for the proposed GAN structure on Vimeo-90K.}
	\renewcommand\arraystretch{1.5}  
	\resizebox{\linewidth}{!}{
		\begin{tabular}{c|c|c|c|c|c|c}
			\hline
			\multirow{2}{*}{Method} & \multicolumn{2}{c|}{  STSR }  & \multicolumn{2}{c|}{ SSR }& \multicolumn{2}{c}{ TSR }\\
			\cline{2-7}
			& PSNR & SSIM & PSNR & SSIM & PSNR & SSIM \\
			\hline
			only generator      & 31.36 &   0.9002  & 32.36  & 0.9324 & 30.01 & 0.8824\\
			only discriminator 1  & 31.61 & 0.9232  & 32.84 & 0.9378 & 30.23 & 0.8931\\
			only discriminator 2  & 31.69 & 0.9258  & 32.82 & 0.9369 & 30.29 & 0.8942\\
			full GAN  & \textbf{31.86} &\textbf{0.9445} & \textbf{32.95}& \textbf{0.9421} & \textbf{30.38} & \textbf{0.8952}   \\
			\hline
		\end{tabular}
	}
	\label{CGAN}
\end{table}

\section{Conclusion}  

We proposed a GAN to reconstruct HR video frames from LR video frames. Unlike previous video STSR methods, the proposed network uses 3D convolutions and a 3D attention mechanism. The generator consists of three parts: shallow feature extraction, deep feature extraction, and reconstruction. In the deep feature extraction part, a 3D CSA is used to enhance important features. The discriminator uses a two-branch structure to deal with details and motion, which further improves the network performance. Experimental results verify the validity of the proposed method. The gains over prior methods are more prominent in regions with rich texture and large motion. 
\textcolor{black}{
In the future, we plan to enhance the performance of our method by leveraging the distinctive features inherent to large motion. Furthermore, we plan to explore combinations with other techniques, such as optical flow estimation.
}

%
%
%

\vfill

\end{document}